\documentclass[a4paper,fleqn,usenatbib]{mnras}

\usepackage{newtxtext,newtxmath}

\usepackage[T1]{fontenc}
\usepackage{ae,aecompl}

\usepackage{graphicx}	
\usepackage{amsmath}	
\usepackage{amssymb}	

\usepackage[normal]{threeparttable}

\title[Molecular gas in the M31 center]{JCMT Mapping of CO(3-2) in the Circumnuclear Region of M31}

\author[Z. Li et al.]{
Zongnan Li,$^{1,2}$
Zhiyuan Li,$^{1,2}$\thanks{E-mail: lizy@nju.edu.cn, mg1726005@smail.nju.edu.cn}
Ping Zhou,$^{3,1,2}$
Yu Gao,$^{4}$
Xue-Jian Jiang,$^{4}$
and Hui Dong$^{5}$
\\
$^{1}$School of Astronomy and Space Science, Nanjing University, Nanjing 210023, China\\
$^{2}$Key Laboratory of Modern Astronomy and Astrophysics, Nanjing University, Nanjing 210023, China\\
$^{3}$Anton Pannekoek Institute, University of Amsterdam, PO Box 94249, 1090 GE, Amsterdam, The Netherlands\\
$^{4}$Purple Mountain Observatory \& Key Laboratory for Radio Astronomy, Chinese Academy of Sciences, Nanjing 210023, China\\
$^{5}$Instituto de Astrof$\acute{\i}$sica de Andaluc$\acute{\i}$a (CSIC), Glorieta de la Astronom$\acute{a}$ S/N, E-18008 Granada, Spain 0000-0002-6403-6817
}

\date{Accepted XXX. Received YYY; in original form ZZZ}

\pubyear{2018}

\begin{document}
\label{firstpage}
\pagerange{\pageref{firstpage}--\pageref{lastpage}}
\maketitle

\begin{abstract}
We present a survey of CO(3-2) molecular line emission in the circumnuclear region of M31 with the James Clerk Maxwell Telescope (JCMT), aiming to explore the physical conditions of the molecular gas. 
Significant CO(3-2) lines are detected primarily along the so-called {\it nuclear spiral}, out to a projected galactocentric radius of 700 pc at a linear resolution of $\sim$50 pc.
We find that the velocity field of the molecular gas is in rough agreement with that of the ionized gas previously determined from optical observations.
Utilizing existed CO(2-1) and CO(1-0) measurements in selected regions of the nuclear spiral,  
we derive characteristic intensity ratios of CO(3-2)/CO(2-1) and CO(3-2)/CO(1-0), which are both close to unity and are significantly higher than the typical intensity ratios in the disk.   
Such line ratios suggest high kinetic temperatures of the gas, which might be due to strong interstellar shocks prevalent in the circumnuclear region.
\end{abstract}


\begin{keywords}
galaxies: individual (M\,31) -- galaxies: nuclei -- galaxies: ISM -- ISM: molecules
\end{keywords}

\section{Introduction} 
Galactic circumnuclear environments, in which the multi-phase interstellar medium (ISM) and various stellar populations are coupled under the influence of a super-massive black hole (SMBH), are the {\it mecca} for a wide array of astrophysical processes. 
Thanks to its proximity ($D \approx$ 780 kpc, corresponding to 1$\arcsec$ = 3.8 pc; \citealp{2006A&A...459...321}), low line-of-sight extinction \citep[$A_{\rm V} \lesssim1$;][]{Dong et al. 2016}, and appropriate inclination ($\sim$$77^\circ$), M31 provides an important and perhaps unique perspective for studying a galactic circumnuclear environment \citep[see][for a recent observational overview]{Li 2009}. 
Central to this environment is the so-called {\it nuclear spiral}, which manifests itself in optical emission lines with a remarkable filamentary morphology across the central kpc of M31 (e.g., \citealp{Jacoby 1985}; Figure~\label{fig:1}). The nuclear spiral is also revealed in infrared emission \citep{Li 2009} and optical extinction against the bulge starlight \citep{2000MNRAS...312...L29,Dong et al. 2016}, indicating that it is composed of neutral, dusty clouds/steamers with ionized outer layers.  
The ionization/excitation mechanism for the nuclear spiral, however, remains a longstanding puzzle, given the lack of circumnuclear massive stars and the extremely quiescent SMBH \citep{Li 2011}.
That the stellar disk of M31 is probably barred \citep{Athanassoula 2006} offers a plausible formation mechanism for the nuclear spiral. By modeling the bar-driven gas dynamics, \cite{Stark 1994} obtained a reasonable fit to the position-velocity diagram jointly derived from the ionized gas in the nuclear spiral and the neutral gas in the outer disk. A possible recent head-on collision between M31 and a satellite galaxy \citep{Block06} may have produced an inward density wave that further modified both the morphology and kinematics of the nuclear spiral. These ideas, however, remain to be confronted with direct observations of the circumnuclear neutral gas, which dominates the mass budget of the nuclear spiral \citep{Li 2009} and could be a more suitable tracer of the kinematics of the latter.

Despite the anticipated gas supply from the outer disk, attempts to directly detecting the circumnuclear neutral gas in M31 have led to only limited success. To date, there is no unambiguous detection of circumnuclear atomic hydrogen reported. An upper limit of $10^6{\rm~M_\odot}$ was placed on the HI mass in the central 500 pc \citep{Brinks 1984}.
A recent survey by \cite{Braun et al. 2009} revealed HI emission in the central few arcminutes of M31, but much of this emission could arise from the outer warped HI disk seen in projection. 
Similarly, surveys of molecular gas \citep[e.g.][]{Loinard 1999}, including the IRAM 30m CO(1-0) survey \citep{2006A&A...453...459-475} with moderate sensitivity (RMS of $\sim$ 0.3 K km s$^{-1}$), did not yield detection in the central region. 
The first detection of circumnuclear CO emission was reported by \cite{2000MNRAS...312...L29}, based on IRAM 30m observations pointed toward a prominent dusty cloud complex at a projected galactocentric radius of $\sim$1.3$'$ ($\gtrsim$300 pc from the nucleus). 
At a few positions closer to the nucleus, significant CO(2-1) emission was detected in deep IRAM 30m observations (\citealp{Melchior & Combes. 2011}, hereafter MC11; \citealp{Melchior 2016}).
\citet[][hereafter MC13]{2013A&A...558A..33A} presented IRAM 30m CO(2-1) observations covering the southern (and fainter) portion of the inner nuclear spiral, which yielded the first detection of molecular gas within the central 30$\arcsec$. 
More recently, \cite{Melchior 2017} presented PdBI interferometric CO(1-0) observations covering the central 100$\arcsec$ under high resolutions, which allowed them to detect of a molecular cloud at only $2\farcs4$ from the nucleus.

In this work, we present the first survey of CO(3-2) emission lines in the central few hundred parsecs of M31, based on sensitive JCMT observations unbiasedly covering regions on and off the nuclear spiral.
It is expected that CO(3-2) emission is sufficiently strong here, due to the prevalence of interstellar shocks and intense radiation field in the circumnuclear region, which would keep the molecular gas warm. 
The strength of CO(3-2) relative to CO(2-1) and CO(1-0) is a robust diagnostic of the gas kinetic temperature, which in turn can be a good indicator of the physical conditions in the circumnuclear region.
We describe the JCMT observations and data reduction procedure in Section 2. 
Results of the CO(3-2) survey are presented in Section 3, along with a detailed comparison with existing measurements of CO(2-1) and CO(1-0).
We summarize our findings and discuss the implications on the gas properties in Section 4. 

\section{Observations and data reduction}

We carried out the CO(3-2) observations with JCMT, taking advantage of the fast scanning capability of its Heterodyne Array Receiver Program (HARP, with a 4$\times$4 receptors array; \citealp{Buckle 2009}).
A total integration of 25.5 hours were accomplished between October-December 2016, all in band 2 weather. 
We used the position-switch basket weave mode to map the circumnuclear region of M31, centered at 00h:42m:44.35s, +41d:16m:08.6s (J2000). 
A clean background region centered at 00h:44m:44.5s, +41d:05m:25.0s 
was chosen.
This technique minimizes the effect of sky and system uncertainties. The basket weave raster map was obtained by scanning the required field horizontally and perpendicularly with the 14 working receptors (receptors H13 and H14 were not operational) of HARP arranged in an obliquely way, moving 1/4 array (i.e., 7\farcs3) for each step, with a separation of $\sim30\arcsec$, thus achieving the Nyquist sampling. 
By design, the raster mode resulted in the lowest RMS noise for the inner $2^{\prime}\times2^{\prime}$ ($\sim$450 pc $\times$ 450 pc) full Nyquist sampling region, but also covers an under-sampling outer region out to a projected galactocentric radius of $\sim$$3'$ (Figure~\ref{fig:1}). All observations had a bandwidth of 1000 MHz with respect to the central frequency of $\sim$346.16 GHz,
and a spectral resolution of 0.488 MHz, equivalent to 0.423 km s$^{-1}$. 
A log of the observations is given in Table~\ref{tab:obs}.

\begin{table}
	\centering
	\caption{Observation log\label{tab:obs}}
	 \begin{threeparttable}
	 \begin{tabular}{cccc}
	\hline
Obs.Date &Time$^a$ & Subscan$^b$ & Central frequency (MHz)\\
\hline
2016-10-27&285&16& 346142.025\\
2016-11-02 &111&6& 346142.025\\
2016-11-05 &72&5 (3)& 346142.025\\
2016-11-06 &174&10& 346142.025\\
2016-11-13 &135&9 (1)& 346176.629\\
2016-11-16 &291&17& 346176.629\\
2016-11-18 &75&4& 346176.629\\
2016-11-25 &72&4& 346176.629\\
2016-11-26 &36&2 (2)& 346176.629\\
2016-12-17 &36&2 (2)& 346176.629\\
2016-12-25 &255&15 (15)& 346176.629\\
\hline
\end{tabular}
 \begin{tablenotes}
   \item[$a$] Integration time in units of minutes.
   \item[$b$] Number of subscans. Each subscan has an integration time of 17.8 minutes. The number of abandoned subscans is given in the parenthesis.
   \end{tablenotes}
\end{threeparttable}
\end{table}

\begin{figure*}
 \includegraphics[width=15cm]{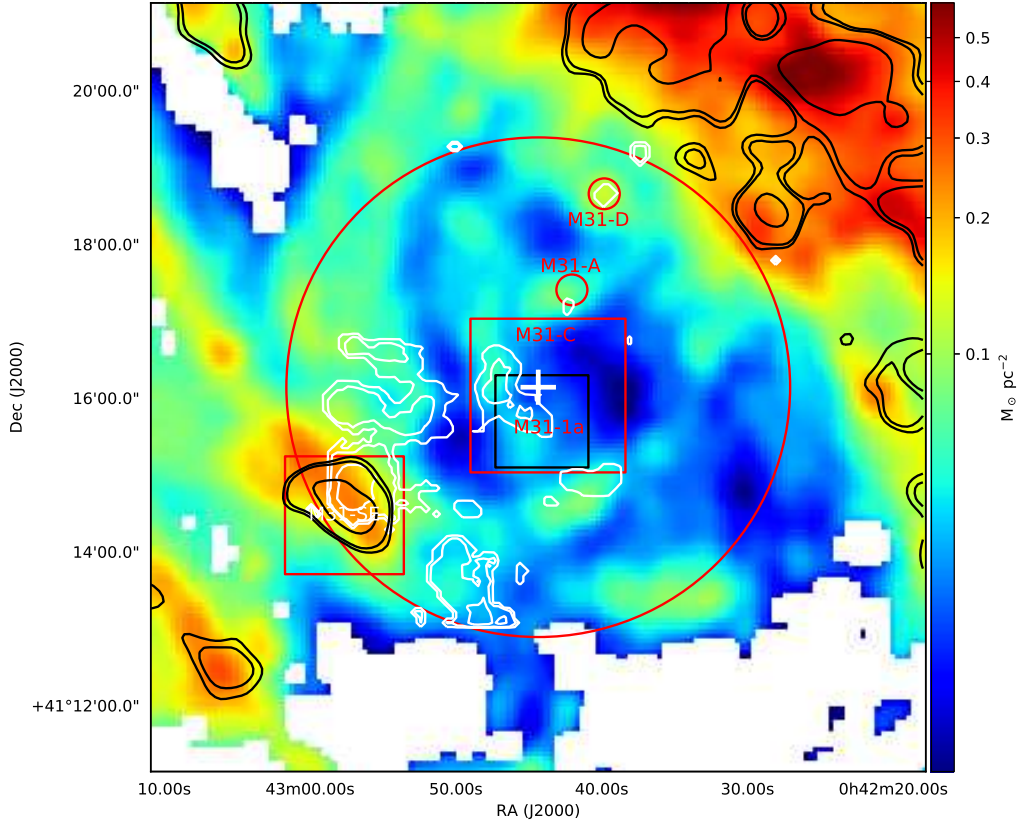}
\caption{Contours of the integrated intensities of CO(3-2) (white, this work) and CO(1-0) (black, \citealp{2006A&A...453...459-475}) overlaid on the dust mass map \citep{2012MNRAS...426...892G} of the circumnulear region of M31. The contour levels are 0.07, 0.35 and 1.75 K km s$^{-1}$ for CO(3-2), 
and 0.9, 1.2 and 3.0 K km s$^{-1}$ for CO(1-0), respectively. 
Both the CO and dust morphologies trace the nuclear spiral. 
The center of M31 is marked with a white cross. The large red circle roughly outlines the field-of-view of the CO(3-2) observations, 
while the large red square indicates the central 2 arcmin $\times$ 2 arcmin region of highest sensitivities, referred to as the M31-C field, which encloses the region observed in CO(2-1) by MC13, referred to as the M31-1a field (shown by the black square). 
The small red square at the southeast corner, partially covering a major filament of the nuclear spiral, is referred to as the M31-SE field. The two small circles labeled with M31A and M31D indicate the positions with CO(1-0) and CO(2-1) observations by MC11. \label{fig:1}}
\end{figure*}

The data were processed with GILDAS\footnote{Continuum and Line Analysis Single-dish Software, http://www. iram.fr/IRAMFR/GILDAS} \citep{Pety 2005}. 
It turned out that the observations taken on 2016-12-17 and 2016-12-25 suffered from wiggle baselines, which might be caused by an instrumental effect, hence this portion of the data was abandoned to avoid systematic problems. 
We also removed a few other questionable subscans to ensure the data quality, and the overall integration time was reduced by 22\%, i.e., 19.3 hours remaining. 
We then used a custom script (Z. Zhang et al. in prep.), originally written for MALATANG, a JCMT Large Program \citep{Tan 2018}, to qualify the spectra, using the ratio of theoretical RMS to measured RMS and the Allan Variance. 
For each selected high quality spectrum, we fitted and subtracted a linear baseline excluding the velocity range of -150 km s$^{-1}$ to 250 km s$^{-1}$ (expected to be occupied by emission lines from M31, see below), corrected for the main beam efficiency of 0.64, and smoothed them to a velocity resolution of 13.5 km s$^{-1}$ for further analysis. For a small fraction ($\sim10\%$) of the spectra, a second order polynomial was employed to characterize the baseline. We produced a position-position-velocity datacube using these spectra, with a spatial resolution of 15$\arcsec$, which is approximately the beam size at 346 GHz. Since the integration time varies across the map, the baseline RMS varies accordingly, with a typical value of $\sim$3.0 mK (in 13.5 km s$^{-1}$ channels) near the field center, gradually increasing to $\sim$20 mK in the outer region (see Figure~\ref{fig:A1} in the Appendix).

Gaussian profiles have been applied to fit tentative CO(3-2) emission lines in the gridded spectra, deriving the line intensity ($I_{\rm CO}$), centroid velocity ($V_0$), FWHM of the line ($V_{\rm FWHM}$) and peak temperature ($T_{\rm peak}$).
We focused on the velocity range of -400 km s$^{-1}$ to 350 km s$^{-1}$, after correcting for M31's systemic velocity of -300 km s$^{-1}$ \citep{van 1969}.
We blindly searched for emission lines across this velocity range with a single Gaussian, 
moving the line center by one channel at each step. 
Only lines with $I_{\rm CO}$ greater than 3\,$\sigma$ were accepted, where $\sigma$ represents the measurement error in $I_{\rm CO}$ from the Gaussian fitting. 
In few cases, two significant lines appear in the spectra, for which we employed double Gaussians to achieve a better fit.
We then produced the integrated intensity map (zero-moment map) and intensity-weighted line-of-sight velocity map (first-moment map), following the method of \cite{Wilson 2012}, which takes into account the varied RMS level across the field-of-view. Briefly, we divided the datacube by the RMS map (Figure~\ref{fig:A1}) to produce a signal-to-noise ratio (S/N) cube, from which pixels with S/N greater than 2 times the local RMS over the velocity range of -150 km s$^{-1}$ to 250 km s$^{-1}$ are identified. 
This velocity range was determined by stacking all gridded spectra and inspecting the width of the resultant broad line. 
We note that this range is consistent with the HI rotation curve \citep{Chemin et al. 2009}, although essentially little HI emission is from the circumnuclear region. 

We have acquired several ancillary datasets to assist our study:
(i) the IRAM-30m CO(1-0) survey by \cite{2006A&A...453...459-475}
with a spatial resolution of 23$\arcsec$ and a typical baseline RMS of 10 mK in 13 km s$^{-1}$ channels in the circumnuclear region, (ii) the IRAM-30m CO(2-1) data from MC13, 
which covers the southern portion of the inner nuclear spiral, with a spatial resolution of 12$\arcsec$ and a baseline RMS of 3.9 mK in 13 km s$^{-1}$ channels, 
and (iii) the dust mass map from \cite{2012MNRAS...426...892G}, 
which was determined by a modified blackbody fit to the $Herschel$ PACS 100, 160 $\mu$m and SPIRE 250 $\mu$m images with a common beam size of 18$\farcs2$. This map is sensitive to a surface dust mass density of $\sim$$0.04{\rm~M_\odot~pc^{-2}}$. 

\section{Results}
\subsection{Morphology and global kinematics}

In Figure~\ref{fig:1}, we compare the morphologies of the circumnuclear molecular gas and dust, by overlaying the CO(3-2) and CO(1-0) integrated intensity contours on the dust mass map, the latter being a good tracer of the nuclear spiral. 
We have convolved the maps of CO(3-2) and dust to match the poorer resolution of CO(1-0) and regridded them to 4$\arcsec$/pixel (same as that of the CO(1-0) map). 
Due to its intrinsic dimness, the CO emission appears patchy as compared to the dust distribution.  
Nevertheless, the distribution of CO(3-2) clearly follows the nuclear spiral.  
This is most clearly seen along a filament in the central $2' \times 2'$ region. Hereafter we refer to this field as M31-C, which encloses the M31-1a field (the $1' \times 1'$ black square in Figure~\ref{fig:1}) observed in CO(2-1) by MC13.
Significant CO(3-2) emission is also evident along an outer filament\footnote{To our knowledge, there exists no specific nomenclature to this filament, which is among the most prominent portion of the nuclear spiral and is part of an off-center ring structure identified by \citet{Block06}.} located at the southeast corner of our field-of-view.
Although the RMS noise in this region (hereafter referred to as the M31-SE field) is substantially higher than in M31-C, the reality of the detected CO(3-2) signals is validated by the presence of spatially-coincident dust and CO(1-0) emission.
Spatially-coincident dust and CO(1-0) emission are also seen at the northwest corner of Figure~\ref{fig:1}, however, this region is beyond the CO(3-2) field-of-view and will not be further discussed.

Except for a few clumps of moderate significance, little CO(3-2) emission can be seen in the western portion of our field-of-view, where the nuclear spiral is also known to be faint in dust emission. 
We further mark in Figure~\ref{fig:1} two positions, M31A and M31D, which have been studied with deep pointed CO(1-0) and CO(2-1) observations by MC11 and also exhibit significant dust emission and extinction in the optical \citep{2000MNRAS...312...L29,Dong et al. 2016}.
Weak CO(3-2) signals are detected at these two positions (see below).

Figure~\ref{fig:2} displays the map of the line-of-sight velocity overlaid with the CO(3-2) integrated intensity contours. The velocity field revealed here is in broad agreement with that traced by the optical emission lines presented in \cite[][Figure 12 therein]{Opitsch 2018}, i.e., redshifted on the eastern side with respect to the minor-axis of M31, consistent with an overall clockwise rotation pattern when viewed from the north pole of the M31 disk. This indicates that the bulk of the molecular gas is kinetically coupled with the ionized gas, which is consistent with the physical picture that the ionized gas is primarily the surface layer of the molecular clouds \citep{Li 2009}. 

\begin{figure}
\includegraphics[width=\columnwidth]{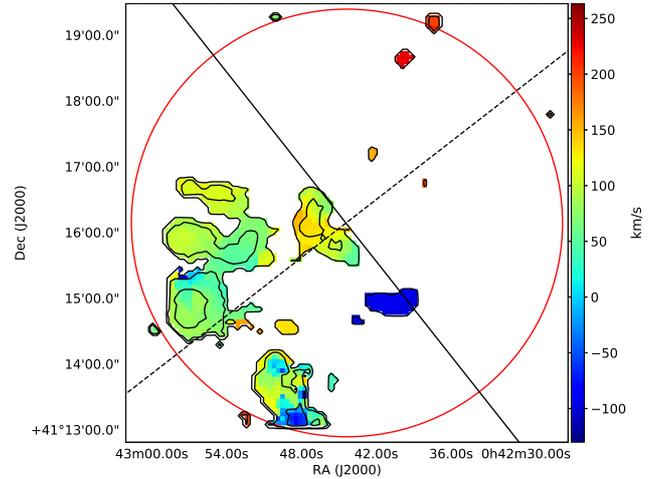}
\caption{The line-of-sight velocity map (the first-moment map) overlaid by the CO(3-2) integrated intensity contours as shown in Figure~\ref{fig:1}. The solid and dashed lines denote the major-axis and minor-axis of M31, respectively. The field-of-view is roughly outlined by the red circle, same as shown in Figure~\ref{fig:1}. The velocities have been corrected for M31's systemic velocity. \label{fig:2}}
\end{figure}

\begin{figure*}
\includegraphics[width=15cm]{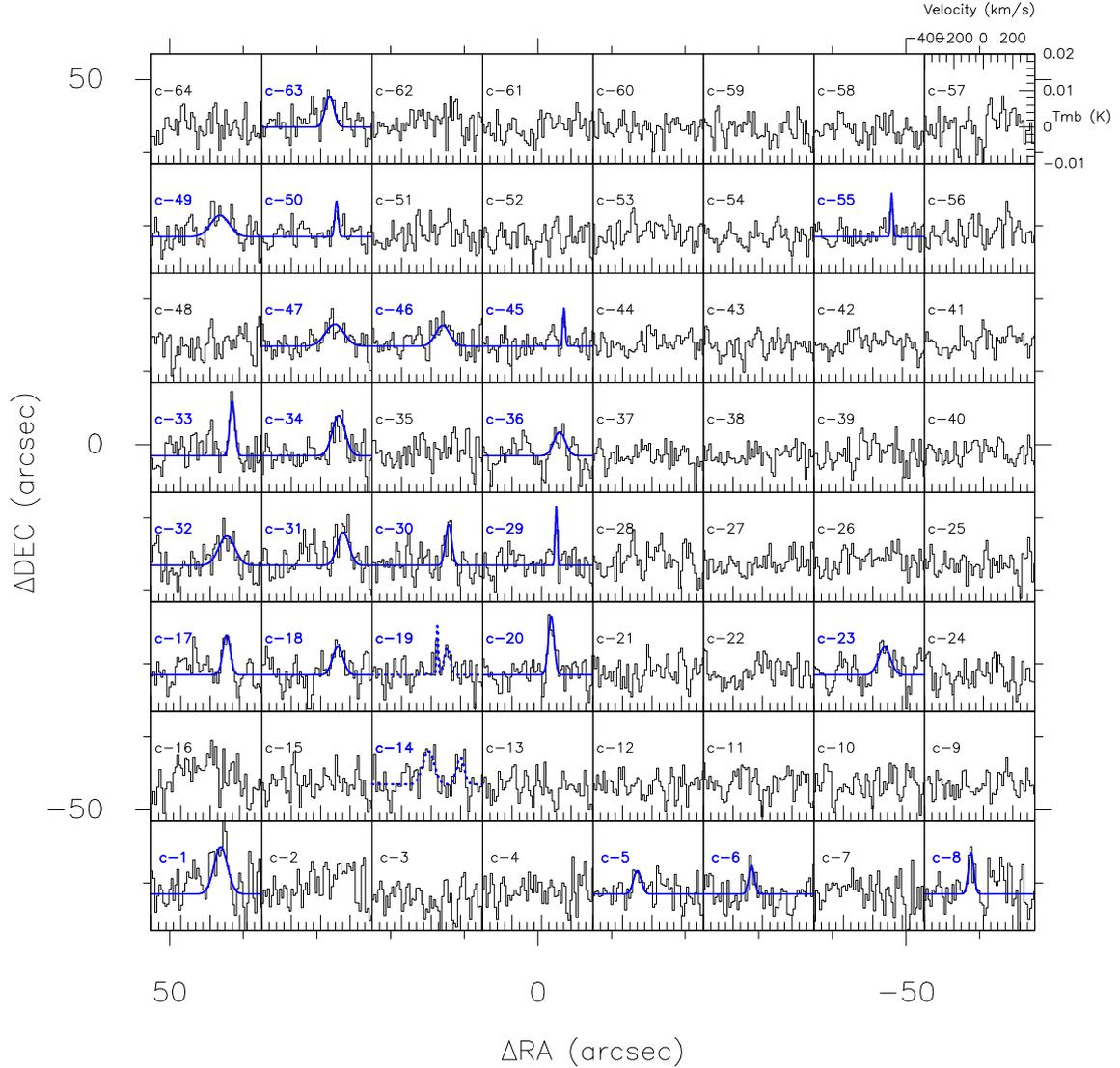}
\caption{The CO(3-2) spectra ({\it black histogram}) of the central 120$\arcsec$ $\times$ 120$\arcsec$ region (M31-C) with a grid size of 15$\arcsec$ (same as the beam size) and a velocity resolution of 13.5 km s$^{-1}$. The coordinate origin is at the M31 center. 
The blue curves are the fitted model with a Gaussian line plus the horizontal baseline. For positions c-14 and c-19, double lines are evident and are thus fitted with two Gaussians, shown as blue dotted curves.
\label{fig:3}}
\end{figure*}

\begin{figure*}
\includegraphics[width=15cm]{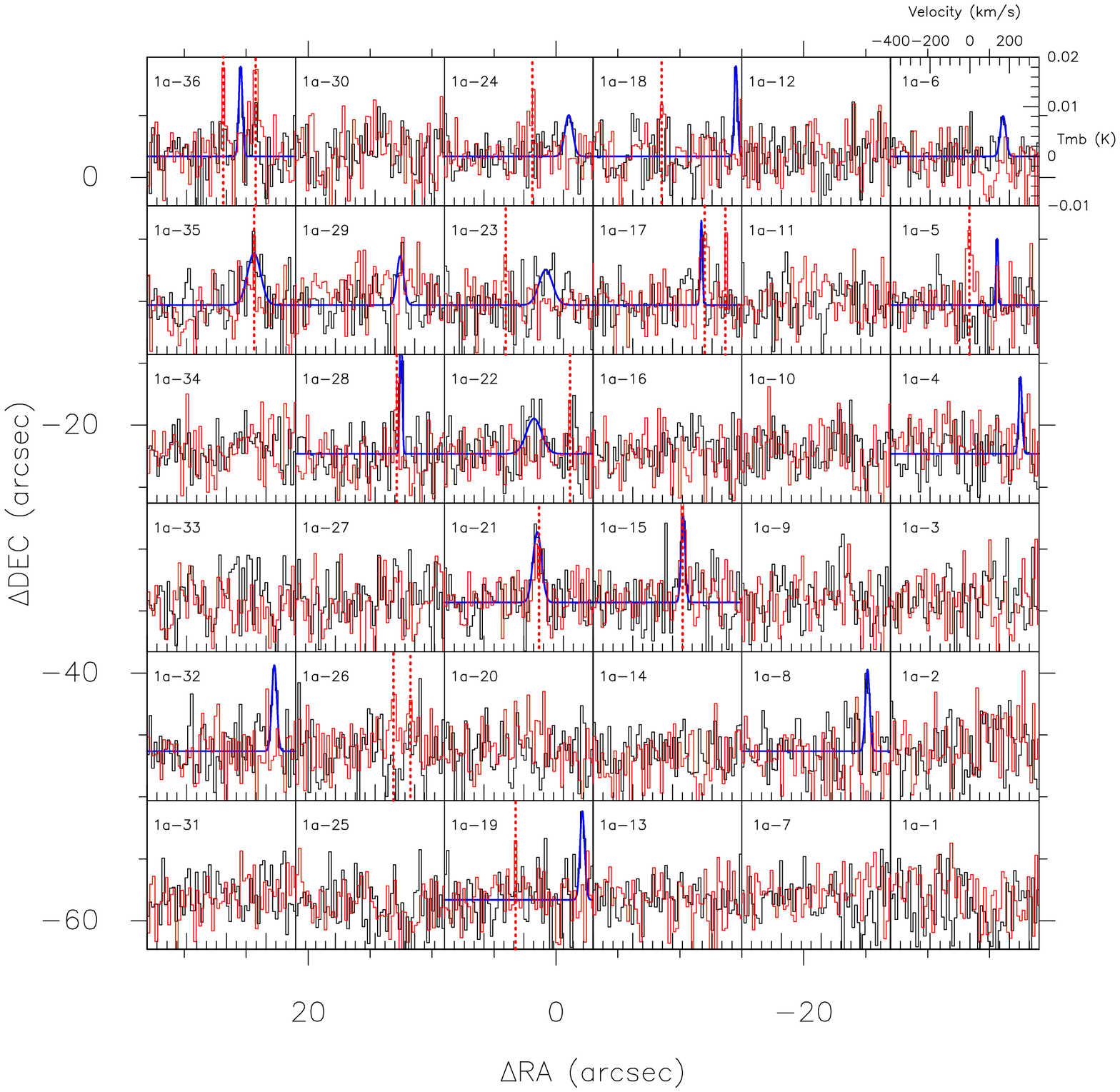}
\caption{Gridded spectra of the M31-1a field, 
with CO(3-2) in black and CO(2-1) in red. The grid size is 12$\arcsec$, same as the beam size of the original CO(2-1) spectra. The M31 center is at the coordinate origin of this map. The blue lines show the fitted Gaussian plus the horizontal baseline. The vertical dotted lines mark the CO(2-1) line centroids of the CO(2-1) reported by MC13.\label{fig:4}}
\end{figure*}


\subsection{Line characteristics and line ratios}

In this section, we describe the line characteristics in the various fields/regions as outlined in Figure~\ref{fig:1}, with a special emphasis on line ratios.
Line ratios play an important role in determining the kinetic temperature and density of molecular clouds. However, since there were no robust detections of CO(1-0) emission in the M31 center, 
the ratios have been inspected only in several positions (MC11). With our CO(3-2) observations and the CO(2-1) and CO(1-0) observations from MC11, MC13 and \cite{2006A&A...453...459-475}, the line ratios can be systematically analyzed for the first time in this region. According to the results of the IRAM-30m extragalactic CO line survey \citep{leroy 2009}, the typical CO(2-1)/CO(1-0) line ratio (hereafter $R_{21}$) of 18 nearby galaxies is $\sim$0.8 in their disks and rising to $\sim$1.3 in their centers. 
Similarly, MC11 found that this ratio in two positions (M31A and M31D) of the nuclear spiral is about unity, whereas \cite{2006A&A...453...459-475} found an average $R_{21}$ $\sim$0.65 in several selected regions on the M31 disk. 
As for the CO(3-2)/CO(1-0) line ratio, $R_{31}$, the typical values are found to be 0.4--0.5 in the disk of the Milky Way \citep{Sanders et al. 1993,Oka et al. 2007}, and similar values in normal spiral galaxies, such as NGC 4254 and NGC 4321 \citep{Wilson 2009}, M51 \citep{Vlahakis et al. 2013}, 
and NGC 628 \citep{Muraoka 2016}. 
\cite{Wilson 2012} reported a mean $R_{31}$ of $0.18\pm0.02$ for 11 nearby galaxies of the JCMT Nearby Galaxies Legacy Survey, and they also obtained a mean CO(3-2)/CO(2-1) line ratio of $0.36\pm0.04$ for 9 galaxies in this survey. 
Moreover, \cite{Leech 2010} found an average $R_{31}$ of 0.47 in a survey of merging sequence of luminous infrared galaxies. 
Comparing these ratios with that of the M31 circumnuclear region could provide us with a deeper insight on the physical conditions of the molecular gas therein. 

\subsubsection{M31-C and M31-1a}

We produced a $8\times8$ gridded spectra map of M31-C with a grid size of 15$\arcsec$, as shown in Figure \ref{fig:3}. 
The typical RMS in these spectra is $\sim$3.5 mK in 13.5 km s$^{-1}$ channels. As shown in this figure, 24 out of 64 positions have CO(3-2) signals greater than 3\,$\sigma$. 
Among them, a blueshifted line is found at positions c-5, c-6, c-8 and c-14, while all other lines are redshifted. 
We note that positions c-14 and c-19 exhibit double components, with a velocity separation of $\sim$220 km s$^{-1}$ and $\sim$70 km s$^{-1}$, respectively.
A number of other positions, though well fitted by a single Gaussian, exhibits a substantial line width ($\gtrsim$50 km s$^{-1}$). 
This suggests that we have integrated the emission from different gas components projected within the same position. 
We note that the relatively broad CO line widths are compatible with those of the ionized gas in the nuclear spiral, which has a mean value of $\sim$90 km s$^{-1}$ under a much finer spatial resolution \citep{Opitsch 2018}. 
The characteristics of the detected CO(3-2) lines are summarized in Table \ref{tab:co3-2}.
We have also averaged the spectra of the other 40 positions; a broad component appeared in the averaged spectrum with an estimated intensity of $0.29\pm0.06$ K km s$^{-1}$. 
The CO(1-0) data \citep{2006A&A...453...459-475}, on the other hand, show no significant signals across the M31-C field; we derived a 3\,$\sigma$ upper limit of 2.7 K km s$^{-1}$ based on the baseline RMS of the averaged CO(1-0) spectrum.

\begin{table}
\centering
\caption{Characteristics of CO(3-2) lines in the M31-C field \label{tab:co3-2}}
\begin{threeparttable}
\begin{tabular}{cccccc}
\hline
Grid$^a$ & $I_{\rm CO(3-2)}$$^b$ & $V_0$$^c$ &$V_{\rm FWHM}$$^d$ & $T_{\rm peak}$ & RMS\\
 & K km s$^{-1}$ & km s$^{-1}$ &km s$^{-1}$ & mK &mK\\
 \hline
c-1 & 1.47 $\pm$  0.30 & 71.1$\pm$11.9&107.7$\pm$23.0&12.8&5.5\\
c-5 & 0.42 $\pm$  0.13 & -100.1$\pm$10.5&61.7$\pm$19.7&6.4&3.0\\
c-6 & 0.37 $\pm$  0.10 & -74.2$\pm$7.8&45.1$\pm$9.8&7.8&3.5\\
c-8 & 0.52 $\pm$  0.17 & -83.6$\pm$7.7&43.5$\pm$16.2&11.2&4.6\\
c-14 & 0.89 $\pm$ 0.17 & -18.0$\pm$8.6&91.4$\pm$19.5&9.2&3.3\\
c-14 & 0.48 $\pm$ 0.13 & 198.1$\pm$9.3&62.0$\pm$13.8&7.3&3.3\\
c-17 & 0.69 $\pm$  0.13 & 110.8$\pm$6.2&59.0$\pm$10.6&11.0&3.5\\
c-18 & 0.70 $\pm$ 0.22 & 117.0$\pm$12.0 & 84.4$\pm$36.1&7.7&4.0\\
c-19 & 0.19 $\pm$ 0.06 & 42.4$\pm$2.4 & 13.5$\pm$392.8 & 13.3&3.0\\
c-19 & 0.37 $\pm$ 0.12 & 108.6$\pm$7.0 & 47.0$\pm$17.2 &7.4&3.0\\
c-20 & 0.79 $\pm$  0.11 & 65.3$\pm$3.6&46.4$\pm$6.7&16.0&3.1\\
c-23 & 0.77 $\pm$  0.21 & 78.7$\pm$16.8&94.7$\pm$27.5&7.6&4.1\\
c-29 & 0.30 $\pm$  0.08 & 99.5$\pm$2.4&17.6$\pm$4.4&16.2&3.1\\
c-30 & 0.60 $\pm$  0.17 & 118.6$\pm$8.2&49.9$\pm$19.6&11.3&3.9\\
c-31 & 0.92 $\pm$  0.21 & 152.7$\pm$11.0&95.0$\pm$23.8&9.1&4.0\\
c-32 & 1.10 $\pm$  0.28 & 112.2$\pm$15.5&129.0$\pm$40.9&8.0&4.2\\
c-33 & 0.66 $\pm$  0.15 & 149.4$\pm$4.6&41.8$\pm$10.6&14.9&4.1\\
c-34 & 1.20 $\pm$  0.22 & 122.5$\pm$9.5&102.4$\pm$20.8&11.0&4.0\\
c-36 & 0.63 $\pm$  0.18 & 122.4$\pm$14.6&91.4$\pm$27.4&6.5&3.6\\
c-45 & 0.22 $\pm$  0.07 & 152.1$\pm$3.4&19.6$\pm$6.1&10.5&2.8\\
c-46 & 0.67 $\pm$  0.18 & 77.7$\pm$16.3&112.6$\pm$27.2&5.6&3.4\\
c-47 & 0.95 $\pm$  0.16 & 95.5$\pm$12.5&148.8$\pm$27.5&6.0&1.7\\
c-49 & 0.92 $\pm$  0.25 & 65.9$\pm$21.7&148.1$\pm$39.4&5.8&3.9\\
c-50 & 0.28 $\pm$  0.07 & 107.6$\pm$3.5&26.5$\pm$8.6&9.8&2.5\\
c-55 & 0.23 $\pm$  0.06 & 127.5$\pm$2.5&18.1$\pm$4.8&11.9&2.6\\
c-63 & 0.64 $\pm$  0.14 & 61.0$\pm$7.8&71.2$\pm$17.8&8.4&3.1\\
\hline
Other & 0.29 $\pm$ 0.06 & 89.7$\pm$26.2&236.1$\pm$48.4&1.2&0.7\\
\hline
\end{tabular}
\begin{tablenotes}
\item[$a$] Same as shown in Figure~\ref{fig:3}. The line in last row is obtained by averaging the other 40 spectra with non-detections. 
\item[$b$]  Integrated intensity, $I_{\rm CO}=\int T_{\rm mb}dv$.
\item[$c$]  Centorid velocity, corrected for M31's systemic velocity.
\item[$d$] The FWHM of the line.
\end{tablenotes}
\end{threeparttable}
\end{table}

As shown in Figure~\ref{fig:1}, the M31-1a field studied by MC13 is entirely embedded in M31-C, permitting a comparison of the CO(2-1) and CO(3-2) lines.
Figure \ref{fig:4} is a comparison of CO(3-2) (shown in black) and CO(2-1) (red) spectra in the M31-1a field with a common velocity resolution of 13 km s$^{-1}$. Since the CO(2-1) data were observed in Wobbler-switching mode and cannot be applied for spatial convolution, we instead regrid our CO(3-2) spectra to match the resolution of 12$\arcsec$ of the CO(2-1) data.  
Table \ref{tab:m31-1a} summarizes the characteristics of CO(3-2) lines in the M31-1a field, along with the CO(2-1) intensities and the line ratios, $R_{32} \equiv I_{\rm CO(3-2)}$/$I_{\rm CO(2-1)}$. 
We caution that considering the difference in the resolutions (15$\arcsec$ versus 12$\arcsec$) and the corresponding beam smearing effect, these ratios can be underestimated by up to a factor of $(15/12)^2 \approx 1.56$.
By averaging the line intensities listed in Table \ref{tab:m31-1a}, we derived a mean intensity of $0.48\pm0.03$ K km s$^{-1}$ for CO(3-2) and $0.52\pm0.03$ K km s$^{-1}$ for CO(2-1), which lead to an average $R_{32} = 0.92\pm0.08$. This exercise has the caveat of neglecting the velocity difference in the two lines. 
Therefore, we also estimated the average line ratio by using only the four positions with good agreements in both line centroid and line width (reported in italic in Table \ref{tab:m31-1a}), which results in $R_{32} = 0.77 \pm 0.27$. 

\begin{table*}
\centering
\caption{Characteristics of CO(3-2) and CO(2-1) lines of the M31-1a field\label{tab:m31-1a}}
 \begin{threeparttable}
\begin{tabular}{ccccccccc}
\hline
 Grid$^a$  &   $I_{\rm CO(3-2)}$ &  $V_0$ & $V_{\rm FWH M}$ &  ${\rm T_{peak}}$ & RMS & $I_{\rm CO(2-1)}$$^b$ &  $\Delta_V$$^c$ & $R_{32}$$^d$ \\
  &  K km s$^{-1}$ &  km s$^{-1}$ & km s$^{-1}$ &  mK & mK & K km s$^{-1}$ &  (km s$^{-1}$)&  \\
\hline
1a-4 & 0.33 $\pm$  0.11 & 255.5$\pm$3.3&20.0$\pm$7.3&15.5&4.3&(0.34)&-&-\\
1a-5 & 0.18 $\pm$  0.06 & 137.9$\pm$4.3&13.0$\pm$7.7&13.4&4.3&0.51 $\pm$ 0.12&$136.2\pm5.9$&$0.35\pm0.14$\\
1a-6 & 0.29 $\pm$  0.09 & 167.8$\pm$5.2&33.2$\pm$11.1&8.2&2.9&(0.60)&-&-\\
1a-8 & 0.46 $\pm$  0.12 & 234.0$\pm$3.6&26.0$\pm$7.5&16.4&3.9&(0.44)&-&-\\
{\it 1a-15} & 0.46 $\pm$  0.15 & 53.7$\pm$3.6&24.2$\pm$10.3&17.8&4.9&0.60 $\pm$ 0.08&$27.5\pm10.3$&$\mathit{0.77\pm0.27}$\\
{\it 1a-17} & 0.25 $\pm$  0.07 & 145.7$\pm$5.4&13.6$\pm$69.7&17.0&3.4&0.76 $\pm$ 0.13&$16.9\pm6.7$&$\mathit{0.33\pm0.11}$\\
 &  &&&&& 0.26 $\pm$ 0.08 &$121.6\pm6.7$&$0.96\pm0.40$\\
1a-18 & 0.38 $\pm$  0.11 & 320.0$\pm$3.1&19.8$\pm$5.4&18.2&4.5&0.35 $\pm$ 0.09&$375.0\pm5.1$&$1.09\pm0.42$\\
1a-19 & 0.53 $\pm$  0.17 & 295.4$\pm$3.9&27.9$\pm$12.4&17.9&4.9&0.13 $\pm$ 0.05&$336.8\pm5.6$&$4.08\pm2.04$\\
{\it 1a-21} & 0.77 $\pm$  0.13 & 67.8$\pm$4.8&51.1$\pm$8.8&14.2&3.5&0.50 $\pm$ 0.13&$8.2\pm8.4$&$\mathit{1.54\pm0.48}$\\
1a-22 & 0.68 $\pm$ 0.21 & 51.4$\pm$15.9& 89.7$\pm$27.4&7.1&4.5&0.20 $\pm$ 0.07&$181.5\pm16.4$&$3.40\pm1.59$\\
1a-23 & 0.62 $\pm$ 0.20 & 110.8$\pm$14.2&81.7$\pm$24.6&7.1&4.5&0.34 $\pm$ 0.09&$201.2\pm14.8$&$1.82\pm0.76$\\
1a-24 & 0.42 $\pm$  0.12 & 226.8$\pm$7.7&47.2$\pm$14.6&8.3&3.2&0.33 $\pm$ 0.09&$183.4\pm9.2$&$1.27\pm0.50$\\
{\it 1a-28} & 0.45 $\pm$  0.07 & 130.6$\pm$0.8&13.0$\pm$33.1&32.8&3.7&0.36 $\pm$ 0.08&$21.5\pm1.3$&$\mathit{1.25\pm0.34}$\\
1a-29 & 0.43 $\pm$  0.11 & 127.0$\pm$6.5&40.4$\pm$11.0&9.9&3.5&(0.76)&-&-\\
1a-32 & 0.54 $\pm$  0.13 & 241.7$\pm$3.6&29.5$\pm$8.0&17.3&4.6&(0.50)&-&-\\
1a-35 & 0.88 $\pm$  0.22 & 139.4$\pm$9.9&79.1$\pm$25.3&10.4&4.3&0.15 $\pm$ 0.05&$0.3\pm10.4$&$5.87\pm2.45$\\
1a-36 & 0.42 $\pm$  0.11 & 71.6$\pm$3.3&22.0$\pm$5.6&18.1&4.5&0.43 $\pm$ 0.10&$87.2\pm4.3$&$0.98\pm0.34$
\\
 &  &&&&& 0.94 $\pm$ 0.16 &$75.6\pm5.7$& $0.45\pm0.14$\\
\hline
1a-26 & (0.59) &-&-&-&-& 0.48 $\pm$ 0.14 &-&- \\
1a-26 & (0.70) &-&-&-&-& 0.45 $\pm$ 0.16 &-&- \\
\hline
\end{tabular}
\begin{tablenotes}
 \item[$a$] Same as shown in Figure~\ref{fig:4}. The last two rows list the two CO(2-1) lines with no corresponding CO(3-2) detections (3\,$\sigma$ upper limits of the CO(3-2) intensity given in parenthesises).
 \item[$b$] The integrated intensity of CO(2-1). 3\,$\sigma$ upper limits are given in parenthesises for the positions of non-detection.
 \item[$c$] The velocity separation of the two lines.
 \item[$d$] The intensity ratio, $I_{\rm CO(3-2)}$/$I_{\rm CO(2-1)}$. The four values adopted to calculate the mean line ratio are in italic.
\end{tablenotes}
 \end{threeparttable}
\end{table*}

\subsubsection{Other regions}

\begin{figure}
\includegraphics[width=\columnwidth]{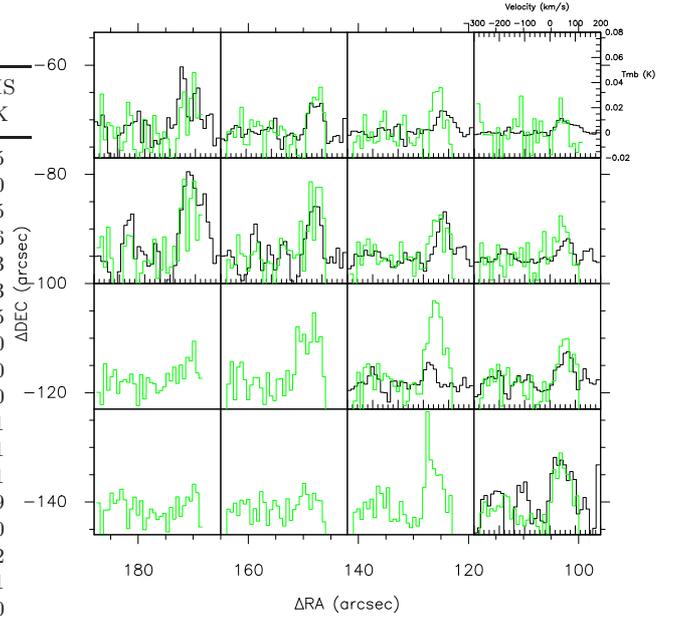}
\caption{Gridded spectra of the M31-SE field. Green for CO(1-0) and black for CO(3-2). The M31 center is at the coordinate origin of this map. The five positions in the lower left corner has no CO(3-2) data due to the limited field-of-view. Both are convolved to a spatial resolution of 23$\arcsec$ and smoothed to a velocity resolution of 13 km s$^{-1}$.\label{fig:5}}
\end{figure}

The CO(3-2) and CO(1-0) gridded spectra of the M31-SE field are shown in Figure~\ref{fig:5}. 
Significant CO(3-2) emission is detected in all 11 positions (the other 5 positions are outside the CO(3-2) field-of-view). It can be seen that the two lines are well correlated with each other, exhibiting similar line centroids at most positions. 
We obtained the total line intensities of the 11 positions by averaging the relevant spectra and fitting a single Gaussian, an exercise justified by the coherent line-of-sight velocities at these positions. This results in 2.14 $\pm$ 0.22 K km s$^{-1}$ for CO(3-2) and 2.38 $\pm$ 0.22 K km s$^{-1}$ for CO(1-0), and thus an average $R_{31} = 0.90\pm0.14$. 
This again points to a systematically higher line ratio in the circumnuclear region than in the disk, which has a mean value of $\sim$0.4-0.5 as mentioned above.  

For M31A and M31D, integrated intensities of CO(2-1) and CO(1-0) have been quoted from MC11. 
We found that the two line ratios, $R_{31}$ and $R_{32}$, are $\sim$0.3 in both M31A and M31D, which are relatively low compared to the cases of M31-1a and M31-SE. 
A possible interpretation is that the molecular gas in M31A and M31D are actually located significantly further away from the M31 center, given their projected positions close to the minor-axis. 

The line intensities and line ratios discussed in the above are summarized in Table \ref{tab:ratio}.

\begin{table}
\scriptsize
\caption{Intensity ratios in various regions. \label{tab:ratio}}
\begin{threeparttable}
\begin{tabular}{cccc|cc}
\hline
 Region$^a$ &  $I_{\rm CO(3-2)}$&   $I_{\rm CO(2-1)}$ &   $I_{\rm CO(1-0)}$ 
 &  $R_{31}$$^b$  &  $R_{32}$$^c$ \\
\hline
M31A  & 0.22$\pm$0.05 & 0.88$\pm$0.06 & 0.75$\pm$0.03 &0.29$\pm$0.23&0.25$\pm$0.24\\
M31D &0.88$\pm$0.13&3.45$\pm$0.16&3.43$\pm$0.10&0.26$\pm$0.15&0.26$\pm$0.15\\
M31-SE &2.14$\pm$0.22&-&2.38$\pm$0.22&0.90$\pm$0.13&-\\  
M31-1a&0.48$\pm$0.03&0.52$\pm0.03$&$<$2.7&-&0.92$\pm$0.08\\    
\hline
\end{tabular}
\begin{tablenotes}
\item[$a$] The regions as shown in Figure~\ref{fig:1}. All intensities, in units of K km s$^{-1}$, are measured at the same resolution (24$\arcsec$), except for the CO(3-2) and CO(2-1) intensities of the M31-1a field, which are at their original resolution of 15$\arcsec$ and 12$\arcsec$, respectively.
\item[$b$] The intensity ratio of CO(3-2)/CO(1-0). 
\item[$c$] The intensity ratio of CO(3-2)/CO(2-1).
\end{tablenotes}
\end{threeparttable}
\end{table}

\section{Summary and Discussion}

We have conducted the first sensitive survey of CO(3-2) emission lines within the central few hundred parsecs of M31.  
Our main results include:

\begin{itemize}
\item CO(3-2) lines are detected primarily along the nuclear spiral. The line-of-sight velocity and velocity dispersion of the molecular gas are in rough agreement with those of the ionized gas measured in the optical band. 
\item In various positions, the CO(3-2) lines are found associated with either CO(2-1) or CO(1-0) lines detected in previous work. The estimated line ratios, $R_{32}$ and $R_{31}$, are close to unity and are significantly higher than the typical values found in the disk of normal spiral galaxies.
\end{itemize}

\begin{figure}
\includegraphics[width=\columnwidth]{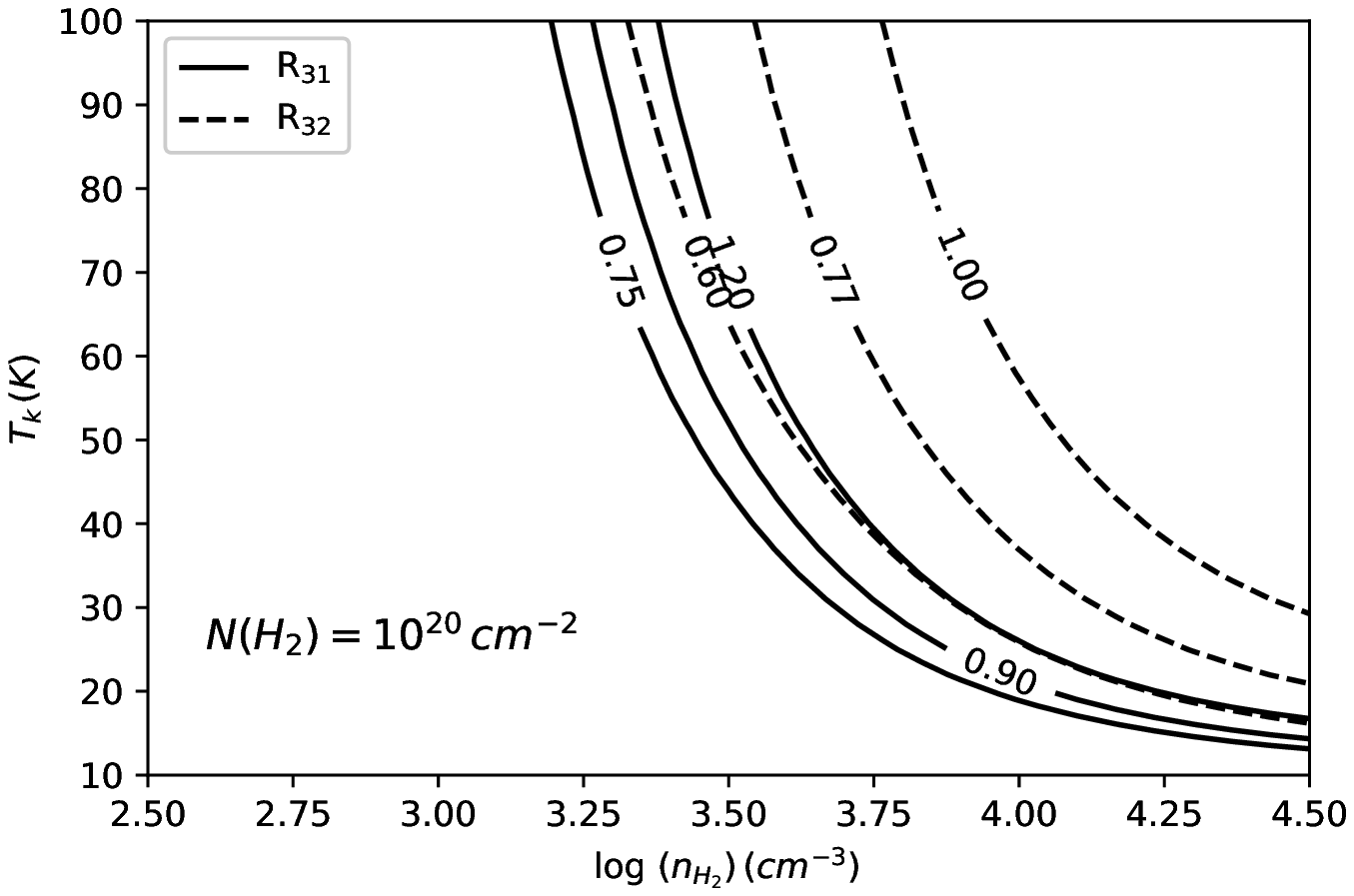}
\includegraphics[width=\columnwidth]{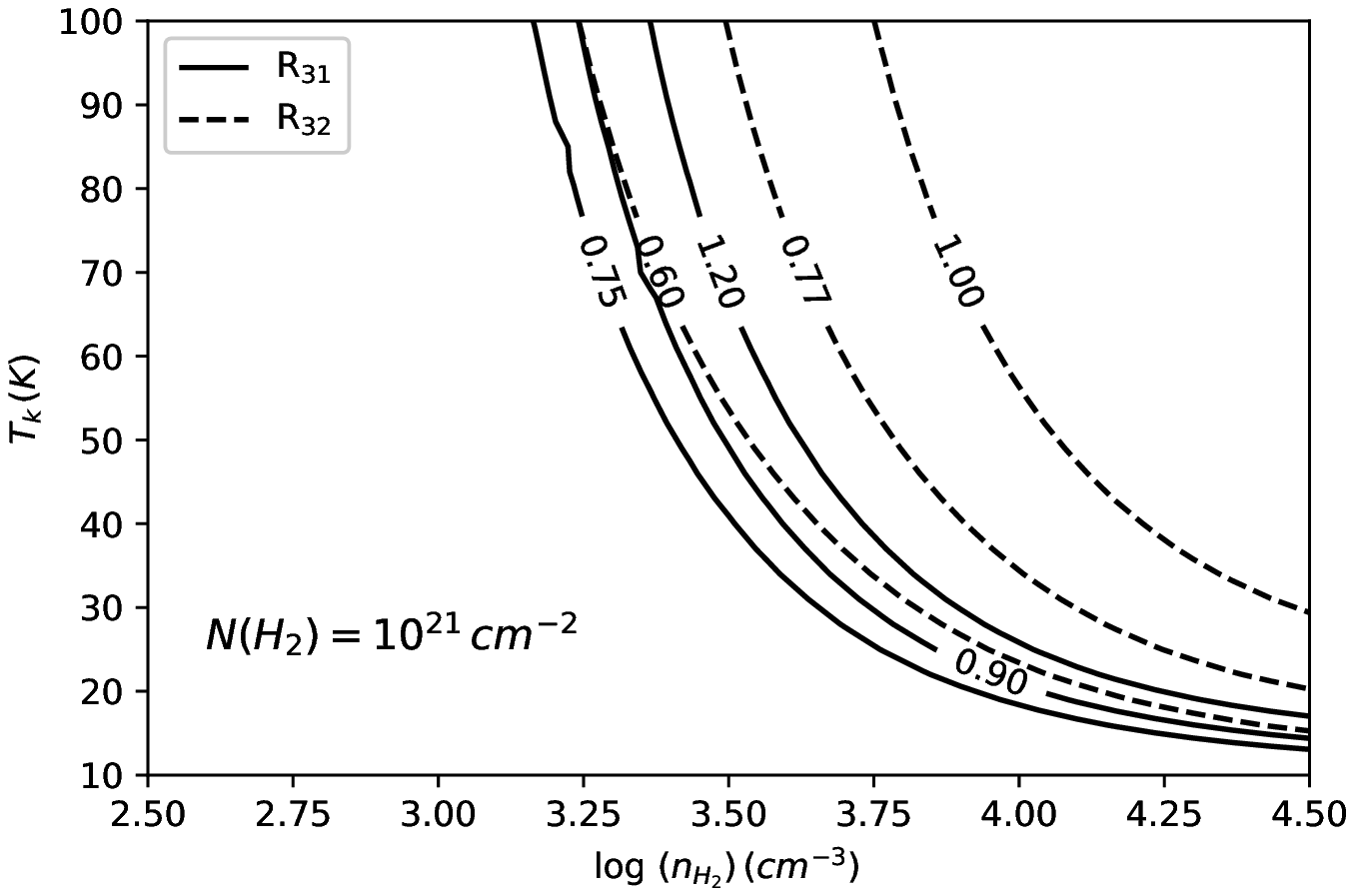}
\caption{Predicted intensity ratios as a function of the kinetic temperature and molecular hydrogen density. The upper and lower panels assume a column density of N(H$_2$)$=10^{20}{\rm~cm^{-2}}$ and $10^{21}{\rm~cm^{-2}}$, respectively. The solid and dashed contours represent ratios of CO(3-2)/CO(1-0) and CO(3-2)/CO(2-1), respectively. \label{fig:6}}
\end{figure}

To constrain the physical conditions of the molecular gas with the high line ratios, we carry out radiation transfer calculations of the line ratios using the RADEX code \citep{van der Tak 2007}, which adopts the large velocity gradient (LVG) approximation \citep[e.g.,][]{Scoville & Solomon 1974}. We assume a typical CO abundance ratio $n_{\rm CO}/n_{\rm H_2}$ = $10^{-4}$ \citep{Bolatto 2013}
, a velocity gradient (dV/dR) of 5 km s$^{-1}$ pc$^{-1}$ and a typical line width $\Delta V$ of 50 km s$^{-1}$ for the LVG model. 
The calculations span kinetic temperature T$_{\rm k}$ = 10--100 K, molecular hydrogen density n(H$_2$) = $10^{2.5}$--10$^{4.5}$ cm$^{-3}$, 
and column density N(H$_2$) = $10^{20}$--10$^{21}$ cm$^{-2}$ (MC13; \citealp{Dong et al. 2016}).   
We present in Figure \ref{fig:6} the LVG results at two fixed values of N(H$_2$): $10^{20}$ (upper panel) and 10$^{21}$ cm$^{-2}$ (lower panel). 
The variation in the line ratios as a function of T$_{\rm k}$ and n(H$_2$) behaves similarly in the two cases. 
To reproduce the observed $R_{32}$ ($\sim$0.8), it requires $T_k \gtrsim 30$ K and n(H$_2$) $\gtrsim 4\times10^{3}$ cm$^{-3}$; the conditions for the observed $R_{31}$ ($\sim$0.90) are similar: $T_k \gtrsim 20$ K and n(H$_2$) $\gtrsim 2\times10^3$ cm$^{-3}$.

High integrated CO line intensity ratios are plausible in the circumnuclear region, which was discussed by \cite{Oka et al. 2012} in the case of the Galactic center. They found a typical CO(3-2)/CO(1-0) ratio of $\sim$ 0.7 in the so-called Central Molecular Zone (as high as $\gtrsim1.5$ at some positions), as compared to the typical ratio of $\sim$0.4 in the Galactic disk. 
This is comparable to $R_{31} \sim 0.90$ found in the M31-SE field. 
\cite{Oka et al. 2012} applied the LVG model to fit the overall intensities and line ratios, obtaining a best-fit with n(H$_2$) $\approx 10^{3.5}$ cm$^{-3}$ and T$_k \approx$ 40 K, which is also comparable to our results. 
\cite{Oka et al. 2012} interpreted the high intensity ratios as due to molecular gas shocked-heated by supernovae and/or winds from massive stars in the Galactic center. 
Similarly, \cite{Ao 2013} derived gas kinetic temperatures of $\sim$50-100 K for dense molecular clouds in the Galactic center. 
They suggested that the high temperatures were caused by turbulent energy dissipation and/or cosmic-rays instead of photons. 
These scenarios are most likely also relevant to the circumnuclear region of M31, in view of the current absence of either an AGN or massive stars \citep{Li 2009}. 
We defer a more quantitative study of the gas excitation in the nuclear spiral to future work.

Finally, we provide an estimate to the total molecular gas mass in the M31-C field, i.e., the central $\sim$450 pc$\times$450 pc region.  
To do so, we assume that $R_{31} \approx 0.9$ as derived from M31-SE also applies for the central field, and adopt a canonical scaling factor $X_{\rm CO} = N_{\rm H_2} /I_{\rm CO(1-0)} = 2 \times 10^{20}{\rm~cm^{-2}}({\rm K~km~s^{-1}})^{-1}$\citep{1988A&A...207...1,Dame 2001}. 
Thus we derive a molecular gas mass $\sim$3.7 $\times 10^5$ M$_\odot$ by adding the 24 detections listed in Table \ref{tab:co3-2} and also accounting for the signal from stacking the 40 positions with non-detections.  
This is to be contrasted with the mass of a few $10^{7}$ M$_\odot$ in the Central Molecular Zone, which occupies a similar physical region in the Galactic center. 
We also estimate the molecular gas-to-dust mass ratio in the circumnuclear region, assisted with the dust surface density ($\Sigma_{\rm dust}$) map from \cite{2012MNRAS...426...892G} as shown in Figure~\ref{fig:1}. 
For the M31-SE field, we directly apply the measured CO(1-0) intensities and the aforementioned $X_{\rm CO}$ factor to determine the surface mass density of molecular hydrogen ($\Sigma_{\rm H_2}$).  
The ratio $\Sigma_{\rm H_2}/\Sigma_{\rm dust}$ ranges from $\sim$25--125, with a mean value of $39.8\pm8.0$. 
This is to be contrasted with the gas-to-dust mass of $38.5\pm5.9$ for the central kpc of M31 obtained by \cite{Draine et al. 2014}, which was based on essentially the same CO(1-0) and dust maps as used here, but also on the HI 21\,cm image of \cite{Braun et al. 2009}.
We further determine the mean molecular gas-to-dust mass ratio in the M31-C field, with the additional assumption of $R_{31} = 0.9$. This results in $\Sigma_{\rm H_2}/\Sigma_{\rm dust}$ = $28.2\pm1.2$. 
It is noteworthy that in the central region, the fraction of atomic-to-total gas $\Sigma_{\rm HI}/\Sigma_{\rm gas}$ is approximately 20\%-40\% according to \cite[][Figure 11 therein]{2006A&A...453...459-475}, 
hence the total gas-to-dust ratio may be close to 50. 
We may also be overestimating $\Sigma_{\rm H_2}$, since \cite{Leroy 2011} find that X$_{\rm CO}$ in the inner 2 kpc of M31 is lower by about a factor two. In any case, the gas-to-dust ratio in the central kpc is significantly smaller than the canonical value of 100 and also smaller than that of the M31 disk ($\sim$100 at 10 kpc, \citealp{Draine et al. 2014}), which might be understood as an increasing metallicity toward the galactic center, as claimed by \cite{Giannetti 2017} in the case of the Milky Way.

\section*{Acknowledgments}
This work is supported by National Key Research and Development Program of China (2017YFA0402703) and National Science Foundation of China (grant 11473010 and 11873028). We are grateful to Mark Goski for his strong technical support to our JCMT observations. We owe our thanks to Anne-Laure Melchior for kindly providing us with the CO(2-1) spectra, to Karl Schuster for providing us with the CO(1-0) data, and to Matthew Smith and Junhao Liu for helpful discussions.


\clearpage

\appendix
\section{Additional data information}

Additional data information are given here. Figure~\ref{fig:A1} shows a map of the baseline RMS of the entire field-of-view, overlaid with CO(3-2) and CO(1-0) intensity contours.  Figure~\ref{fig:A2} shows gridded spectra of the central $2'\times2'$ region in the original half-beam, which can be used as a reference to the identification of the lines in this region.

\setcounter{figure}{0}
\renewcommand\thefigure{A\arabic{figure}}
\begin{figure}
\includegraphics[width=\columnwidth]{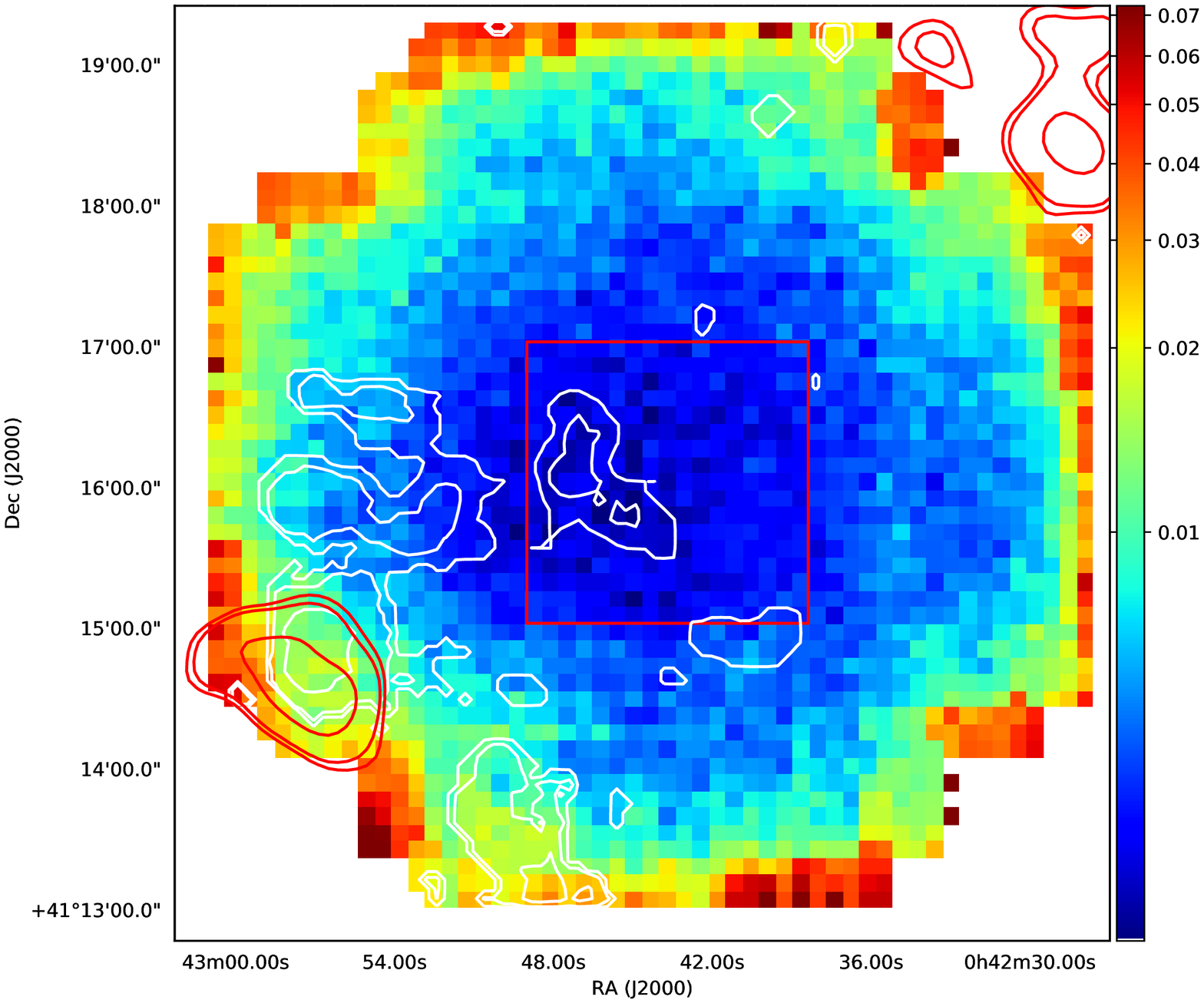}
\caption{The RMS noise of the entire field, overlaid with the CO(3-2) (white) and CO(1-0) (red) contours same as shown in Figure~\ref{fig:1}. \label{fig:A1}}
\end{figure}

\begin{figure*}
\includegraphics[width=15cm]{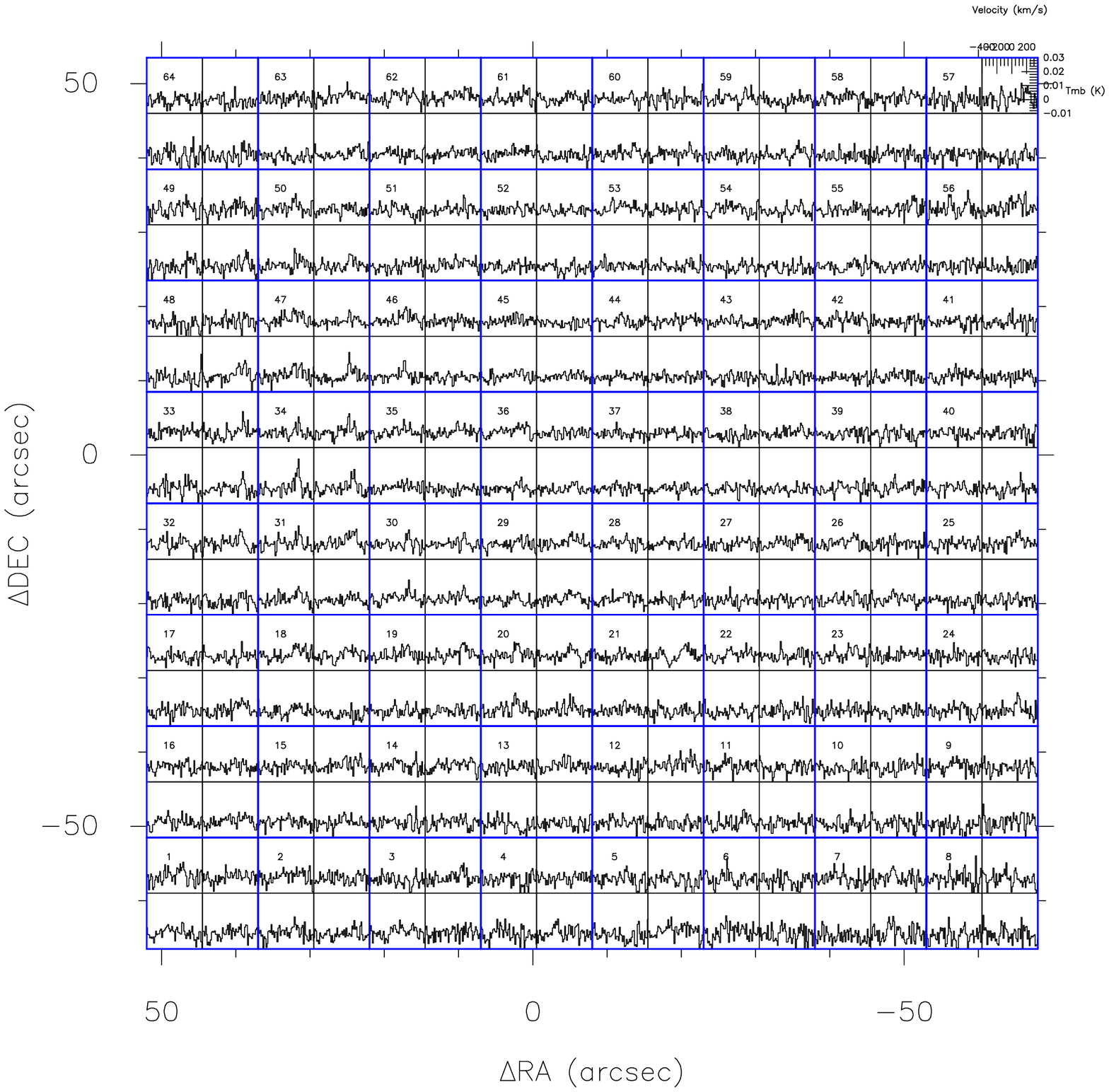}
\caption{The half-beam gridded spectra of the M31-C field. The velocity range of each grid is from -400 to 350 km s$^{-1}$, with a velocity resolution of 13.5 km s$^{-1}$, same as in Figure \ref{fig:3}. T$_{\rm mb}$ ranges from -0.01 to 0.03 K. The numbers refer to the grids as shown in Figure \ref{fig:3}. \label{fig:A2}}
\end{figure*}




\bsp	
\label{lastpage}
\end{document}